# Complete mode conversion for elastic waves reflected by elastic metamaterial slab with double hexapole resonances

Di Liu, Wenjie Yu, Qiujiao Du, Fengming Liu and Pai Peng*

School of Mathematics and Physics, China University of Geosciences, Wuhan 430074, Hubei, China

## Abstract

In this study, we investigate the phenomenon of mode conversion in elastic bulk waves using coupled hexapole resonances. A metamaterial slab is proposed enabling the complete conversion between longitudinal and transverse modes. Each unit of the elastic metamaterial slab comprises a pair of scatterers, and their relative direction is oriented at an oblique angle. The interaction between the coupled hexapoles and the background results in oblique displacements, which are responsible for the mode conversion. Moreover, this conversion exhibits a broader frequency range compared to the quadrupole resonance. This innovative design significantly broadens the range of possibilities for developing mode-converting metamaterials.

paipeng@cug.edu.cn

The elastic bulk waves can be divided into shear waves and longitudinal waves (so called S waves and P waves) according to the relationship between the direction of vibration and the direction of propagation.[1] The mode conversion of elastic waves is related to the coupling between these two modes. In some cases, such as collimation and non-destructive testing applications, the mode conversion effect needs to be minimum[2-4], but in underwater sound absorption, ultrasonic imaging, etc., the mode conversion efficiency needs to be as high as possible[4-7]. In recent years, the rise of elastic wave metamaterials has provided a new method for wave manipulation[8-19] especially for mode conversion.[20-29] Different from converting by inclined incident[30] or scatter[31], the use of metamaterials can solve some limitations of these traditional methods. Kim et al proposed elastic wave metasurface[22] to achieve complete mode conversion. They used a strip unit to form a metasurface to realize the phase shift of $2\pi$ range with its length variation, so as to realize the conversion of longitudinal wave incident at a certain range of angle into the reflected transverse wave. Except through designing phase gradients, mode-coupled layer composed of anisotropic structure[28] proposed by Kim 's group solves the problem of complete converting P to S wave and limited incidence angle. It realizes the efficient conversion of normally incident longitudinal wave into the transmitted transverse wave. Recently, Wang et al.[25, 27] use the anisotropic dipolar resonances to generate sufficiently strong oblique displacements and consequently couple the P and S waves. With the P wave's normal incidence, the reflected waves are modulated to be pure S waves within a space that orders smaller than the working wavelength.

In this work, we propose a design to efficiently realize the complete mode conversion utilizing double-scatterer structure with periodic arrangement. The efficient conversion is realized by the oblique displacement of the boundary caused by the strong coupling of the two hexapoles instead of a single dipole. As long as the frequency of the normally incident wave can inspire the hexapole resonances of the scatterers, the coupling of longitudinal and transverse waves occurs, resulting mode conversion in reflected waves, even complete conversion can be reached after parameters' designing.

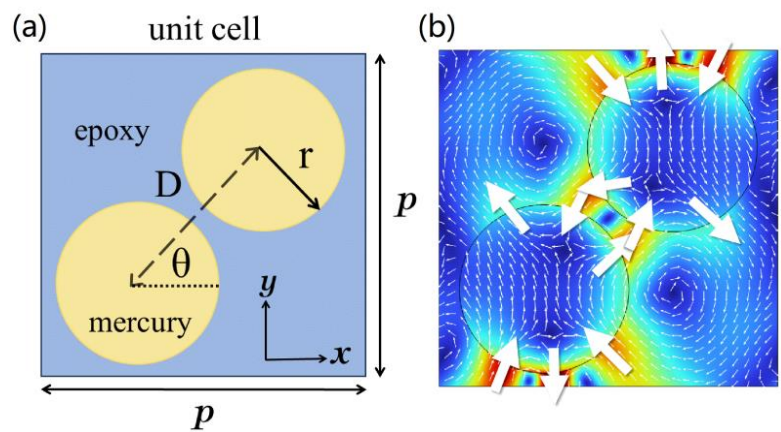

Fig. 1 (a)Schematic illustration of the unit cell. (b)The velocity field distributions for the eigenmode of coupled hexapoles.

The proposed unit cell, as shown in Fig. 1(a), contains two cylindrical scatterers filled with mercury, and they are embedded in epoxy matrix. These two different materials are presented in schematic illustration by yellow and blue colors respectively. The scatterer's radius is $r = 0.2505p$, and the distance between the two cylinders' center is $D = 0.4\sqrt{2}p$, where $p$ is the period constant and also the slab thickness. The two scatterers are relatively oblique with an inclined angle $\theta = 0.267\pi$. The used material parameters are: $\rho_m = 13590 \text{ kg m}^{-3}$, $c_m = 1451 \text{ m s}^{-1}$ for mercury, $\rho_e = 1180 \text{ kg m}^{-3}$, $c_{P,e} = 2540 \text{ m s}^{-1}$, and $c_{S,e} = 1160 \text{ m s}^{-1}$ for epoxy, where $\rho$ is the mass density, and $c_P$ and $c_S$ are speeds for P and S waves, respectively.

The numerical simulation in this research is calculated using the commercial program COMSOL Multiphysics. Bloch boundary conditions are adopted for the left and right boundaries of the unit cell, and the top and bottom boundaries are set as free boundaries, which means the cells are arranged along the x-axis while their top and bottom boundaries are unbound. The intrinsic velocity field of the unit cell is shown in Fig. 1(b), which stands for the coupling of two hexapoles at the normalized frequency $f_1 = 0.4280 f_0$ (with $f_0 = c_{S,e}/p$). The white arrow indicates the velocity direction. Each scatterer of the unit is excited for hexapole resonance, and the two are strongly coupled. The key to realize the mode conversion of elastic wave in this work is to create interaction between the resonators and the background to realize oblique displacements in the matrix. Because of the symmetry, a single scatterer unit does not produce a net oblique displacement on the boundary, whereas this is possible if the double-scatterer structure is used especially when two scatterers are placed at a relative inclined angle.

We consider an elastic metamaterial slab composed of periodic arrangement of units near the surface of a semi-infinite epoxy background, as shown in Fig. 2(a). We do this by using periodic conditions at the left and right boundaries of the calculation region. The boundary load that generates the P wave is placed at a certain distance from the upper free boundary, and such normally incident wave can be regarded as a plane wave.

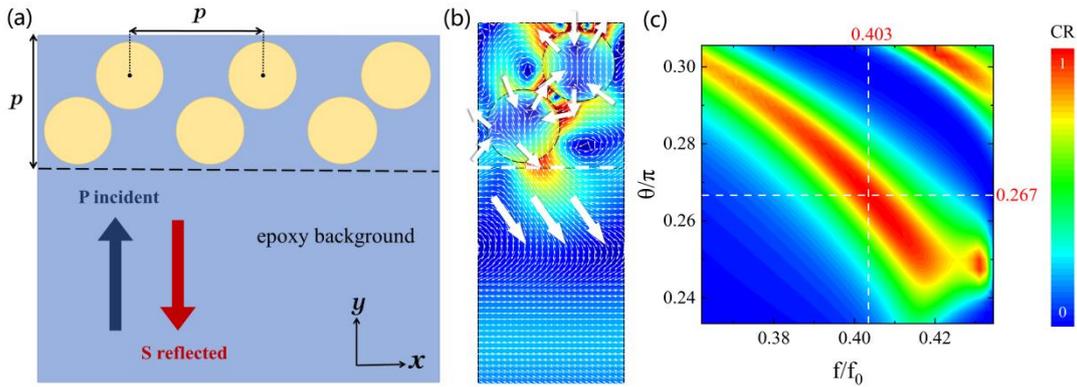

Fig. 2 (a)The schematic illustration of the calculation region. The periodic boundary conditions are applied to the left and right sides. The free and absorbing boundaries are applied to the top and bottom sides, respectively. (b)The velocity field distributions when the complete conversion occurs. The 12 white arrows present the hexapole resonances of the scatterers. (c)Simulation result of the conversion rate varies with $\theta$ and frequency. The crossover point of the white dashed lines indicates the occurrence of the complete conversion.

When the two scatterers are placed at an angle $\theta$ to each other (with $\theta \neq 0$ and $\theta \neq 0.5\pi$), the symmetry of the system is broken. This causes the coupling of transverse and longitudinal waves, which eventually produces the reflected wave of the system containing the S wave component. The conversion rate (CR) between the incident P wave and the reflected S wave is defined by the ratio of their energy fluxes along the vertical direction. We use P-waves of different frequencies to excite hexpole resonances of double scatterers, and different conversion rates can be calculated. The velocity field diagram corresponding to the occurrence of complete conversion (CR=1) is shown in Fig. 2(b) at the frequency of $0.403 f_0$, and twelve white arrows represent the mainly deformation of scatterers. Since the velocity field of the metamaterial part is very similar to the intrinsic vibration mode in Fig. 1(b), we attribute the mode conversion to the excited coupled hexapoles. The coupled hexapole resonances bring about the oblique vibrations to this system. These oblique vibrations should be coupled vertical and horizontal vibrations, so the coupled hexapoles induce vibrations, whose vertical component could come from the incident P wave, and the remaining horizontal component would generate reflect S waves. From the Fig. 2(b), the transverse velocity is also contained in the solid background, indicating that the reflected wave contains S wave.

In addition, we further calculated the ratio of transverse displacement $|u_x|$ and longitudinal displacement $|u_y|$ on the boundary (denoted by the black dashed line in Fig. 2(a) between the elastic metamaterial slab and the solid background, and we can get that $|u_x|/|u_y| \approx 1.42$, which is in accordance with the reference[25].

When inclined angle $\theta$ is changed, the working frequency and maximum conversion rate of the metamaterial slab changes accordingly. The variation of CR with $\theta$ and frequency of the incident P wave is shown in Fig. 2(c). The conversion rate can reach to 100% when $\theta = 0.267\pi$ and $f_1' = 0.403 f_0$ as shown in Fig. 2(c) by the crossed white dashed line. Actually, multiple frequencies near the eigenfrequency $f_1 = 0.4280 f_0$ can excite the hexapole resonances of the scatterers, and brings mode conversion when $\theta$ is appropriate. The $f_1'$ is close to the eigenfrequency, and the interaction between the lower scatterer and the solid background could cause the frequency difference between them.

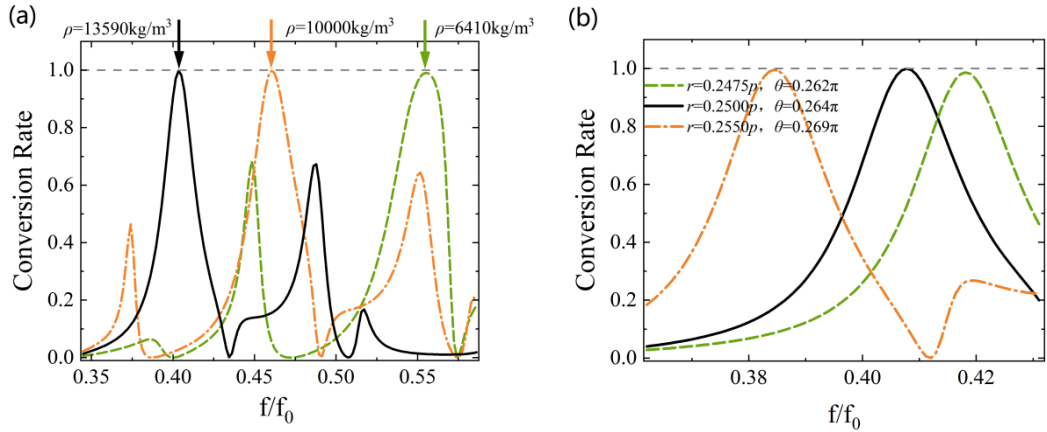

Fig. 3 (a)The CRs vary with frequency for different density of the scatterers' filling liquid. As the density increases, the frequency at which the complete conversion occurs decreases. (b)The CRs vary with frequency for different geometric parameters when mercury is used as the filling liquid. Several models that completely convert P waves at different frequencies can be found.

The efficient working frequency of mode conversion varies with material parameters and geometric parameters. We calculated the conversion rates of structures with different densities of the scatterers' filling liquid, as shown in Fig. 3(a). The $\theta$ is fixed at $0.267\pi$. The complete mode conversion can be obtained at different liquid densities, and the working frequency decreases as the density increases. The solid black line indicates that the filling liquid is mercury. The CRs as a function of frequency under different scatterers' radius $r$ and angle $\theta$ is shown in Fig. 3(b). In order to obtain mode conversion under the lowest possible frequency conditions, we chose mercury as the liquid material. When the scatterers' radius $r$ increases, the high efficient conversion point moves to a lower frequency, which means the double scatterers with larger radii act at lower excitation frequency. If $\theta$ is adjusted slightly at the same time, multiple models that can completely convert P waves at different frequencies can be obtained. This is because the eigenfrequency of each intrinsic vibration mode decreases as the mass of the scatterer increases. If we increase the mass of the scatterer by increasing the density or radius, the working frequency can be further reduced.

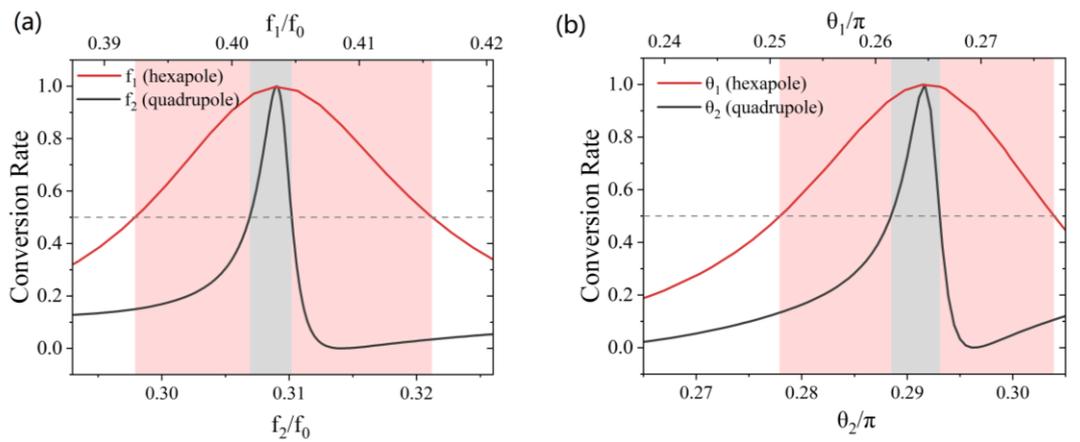

Fig. 4 (a)The curves of CR with frequency. (b)The CRs vary with $\theta$ at their respective excitation frequencies. For both (a) and (b), the black and red lines indicate the coupled quadrupole resonances and coupled hexpole resonances, respectively, that are utilized to convert modes.

Compared with the previous work using coupled quadrupole resonance to realize mode conversion[32], the high conversion region in this work is less sensitive to frequencies, as shown in Fig. 4(a). It presents the curve of CR as a function of frequency. For coupled hexpole resonances, the frequency width of CRs above 0.5 (colored background) is significantly wider than that for quadrupole resonances (gray background). Similarly, for the range of geometric parameter $\theta$ that satisfying conversion rates higher than 0.5, coupled hexpole resonances are also better than quadrupole resonances. This brings a wider range of adaptability to the mode conversion device with double-scatterer structure. For example, the coupled hexapole resonances of the double-scatterer structure can achieve a high conversion rate at a wider bandwidth, and the larger range of $\theta$ provides the basis for its operation on different background materials.

In conclusion, we investigate the mode conversion of elastic waves in a reflection system by a double-scatterer structure's coupled hexapoles. The oblique vibrations induced by hexapoles should be coupled vertical and horizontal vibrations, leading to occurrence of mode conversion. The maximum P to S wave conversion rate can reach 100%. Compared to the coupled quadrupoles, it has a larger frequency range and looser geometric parameter's demand. It reduces the production difficulty by eliminating the requirement for anisotropic resonators and broadens the design space by demonstrating other types of resonances out of the dipole.

1) K.F. Graff: Dover Publications. (1991).
2) X. Su, Z. Lu and A.N. Norris: Journal of Applied Physics. **123** [9](2017).
3) A. Climente, D. Torrent and J. Sánchez-Dehesa: Applied Physics Letters. **105** [6](2014).
4) M. Mitra and S. Gopalakrishnan: Smart Materials and Structures. **25** [5](2016)053001.
5) Y. Tian, Y. Shen, D. Rao and W. Xu: Smart Materials and Structures. **28** [7](2019)075038.
6) A.P. Sarvazyan, O.V. Rudenko, S.D. Swanson, J.B. Fowlkes and S.Y. Emelianov: Ultrasound in Medicine & Biology. **24** [9](1998)1419.
7) R.M. Sturm, E.B. Yerkes, J.L. Nicholas, D. Snow-Lisy, D. Diaz Saldano, P.L. Gandor, C.G. Halline, I. Rosoklija, K. Rychlik, E.K. Johnson and E.Y. Cheng: The Journal of Urology. **198** [2](2017)422.
8) X.Z. Zhengyou Liu, Yiwei Mao, Y. Y. Zhu, Zhiyu Yang, C. T. Chan, Ping Sheng*: Science. **289** (2000).
9) Y. Ding, Z. Liu, C. Qiu and J. Shi: Physical Review Letters. **99** [9](2007)093904.
10) N. Fang, D. Xi, J. Xu, M. Ambati, W. Srituravanich, C. Sun and X. Zhang: Nature Materials. **5** [6](2006)452.
11) J. Li and C.T. Chan: Physical review. E, Statistical, nonlinear, and soft matter physics. **70** [5 Pt 2](2004)055602.
12) Y. Wu, Y. Lai and Z.Q. Zhang: Physical Review Letters. **107** [10](2011)105506.
13) D. Beli, M.I.N. Rosa, C. De Marqui and M. Ruzzene: Physical Review Research. **4** [4](2022)043030.
14) W. Jiang, M. Yin, Q. Liao, L. Xie and G. Yin: International Journal of Mechanical Sciences. **190** (2021)106023.
15) Y. Li, H. Li, X. Liu and S. Yan: Journal of Physics D: Applied Physics. **55** [5](2022)055303.
16) Z. Wang, Z. Ma, X. Guo and D. Zhang: Applied Mathematics and Mechanics. **42** [11](2021)1543.
17) Z. Lin, W. Xu, C. Xuan, W. Qi and W. Wang: Journal of Physics D: Applied Physics. **54** [25](2021)255303.
18) B. Djafari-Rouhani, L. Carpentier and Y. Pennec: 2022 IEEE International Ultrasonics Symposium (IUS), 2022, p. 1.
19) X. Cao, C. Jia, H. Miao, G. Kang and C. Zhang: Smart Materials and Structures. **30** [5](2021)055013.
20) J.M. De Ponti, L. Iorio, E. Riva, R. Ardito, F. Braghin and A. Corigliano: Physical Review Applied. **16** [3](2021).
21) Y. Guo, F. Liu, Q. Du and P. Peng: Applied Physics Express. **15** [12](2022)127002.
22) M.S. Kim, W.R. Lee, Y.Y. Kim and J.H. Oh: Applied Physics Letters. **112** [24](2018)241905.
23) X. Li, Y. Chen, X. Zhang and G. Huang: Extreme Mechanics Letters. **39** (2020)100837.
24) Y. Noguchi, T. Yamada, M. Otomori, K. Izui and S. Nishiwaki: Applied Physics Letters. **107** [22](2015).
25) Q. Wang, W. Yu, H. Chang, D. Qiujiao, F. Liu, Z. Liu and P. Peng: Applied Physics Express. **15** [11](2022)117001.
26) X. Yang, T. Wang, Y. Chai and Y. Li: Journal of Physics D: Applied Physics. **55** [3](2021)035302.
27) W. Yu, P. Peng, W. Hu, Q. Du and F. Liu: Applied Physics Express. **16** [1](2023)017001.
28) J.M. Kweun, H.J. Lee, J.H. Oh, H.M. Seung and Y.Y. Kim: Physical Review Letters. **118**

| | |
|---|---|
| | [20](2017)205901. |
| 29) | Y. Tian, Y. Shen, X. Qin and Z. Yu: Applied Physics Letters. **118** [1](2021). |
| 30) | A. Bedford and D. Drumheller: Introduction of Wave Propagation (Wiley, New York, 1994). 151. |
| 31) | J. Miklowitz and R.K. Kaul: Journal of Applied Mechanics. **46** [4](1979)969. |
| 32) | D. Liu, P. Peng, W. Yu, Q. Du and F. Liu: Applied Physics Express. **16** [7](2023)077001. |